\documentclass[superscriptaddress,twocolumn,showpacs,prl]{revtex4}

\usepackage{graphicx}  
\usepackage{times}
\usepackage{bm}
\usepackage{amssymb,amsfonts,amsmath}

\newcommand{\bra}[1]{\left(#1\right)}
\newcommand{\Bra}[1]{\left[#1\right]}

\begin{document}

\title{Drifting solitary waves in a reaction-diffusion medium with differential advection}

\author{Arik Yochelis}
\affiliation{Department of Chemical Engineering, Technion -- Israel Institute of Technology, Haifa 32000, Israel}

\author{Moshe Sheintuch}
\affiliation{Department of Chemical Engineering, Technion -- Israel Institute of Technology, Haifa 32000, Israel}

\received{\today}
\begin{abstract}
Propagation of solitary waves in the presence of autocatalysis, diffusion, and symmetry breaking (differential) advection, is being
studied. The focus is on drifting (propagating with advection) pulses that form via a convective instability at lower reaction rates of the
autocatalytic activator, i.e. the advective flow overcomes the fast excitation and induces a drifting fluid type behavior. Using spatial
dynamics analysis of a minimal case model, we present the properties and the organization of such pulses. The insights underly a general
understanding of localized transport in simple reaction-diffusion-advection models and thus provide a background to potential chemical and
biological applications.
\end{abstract}

\maketitle

Solitary waves are prominent generic solutions to reaction-diffusion (RD) systems and basic to many applied science
disciplines~\cite{excitable}. In one spatial dimension, these spatially localized propagating pulses are qualitatively described by a fast
excitation (leading front) from a rest state followed by a slow recovery (rear front) to the same uniform state~\cite{excitable}. Thus, in
isotropic RD media a single symmetric supra-threshold localized perturbation results in \textit{counter-propagating} pulses or wave
trains~\cite{YKXQG:08}.

However, in chemical and biological media transport can be facilitated by both diffusion and advection and thus excitation properties of
solitary waves can be subjected to convective instabilities~\cite{adv}. Nevertheless, theoretical foundations of solitary waves in the
presence of a symmetry breaking advective transport, i.e. dynamics in a differential reaction-diffusion-advection (RDA) media, are yet to
be established. Among a few reported examples, it was only shown both experimentally and numerically (with no underlying theoretical
basis), that excitable pulses can persist in RDA with a propagation direction against the advective field~\cite{KM:02}, i.e. upstream.

\begin{figure}[tp]
\includegraphics[width=3.1in]{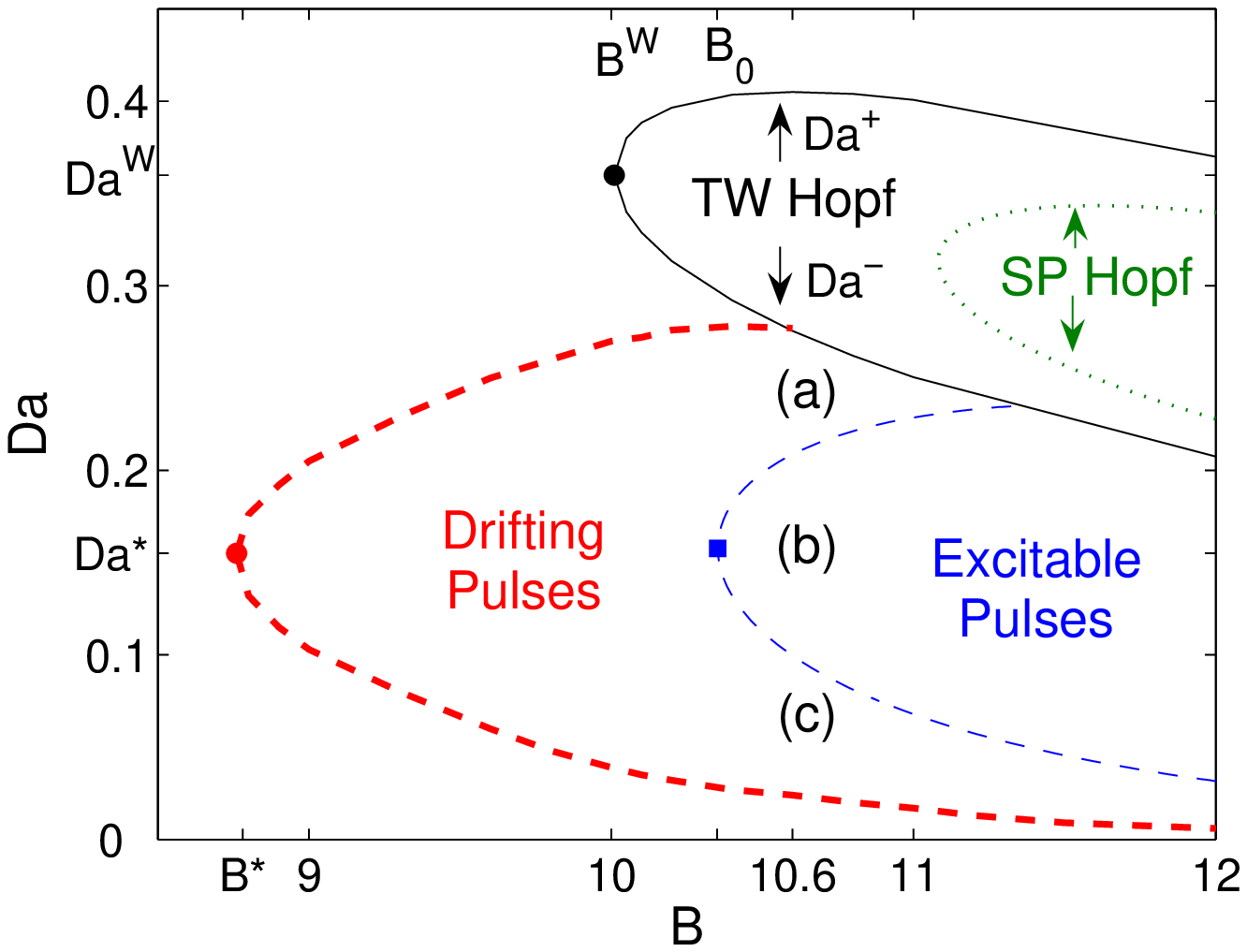}
\includegraphics[width=3.1in]{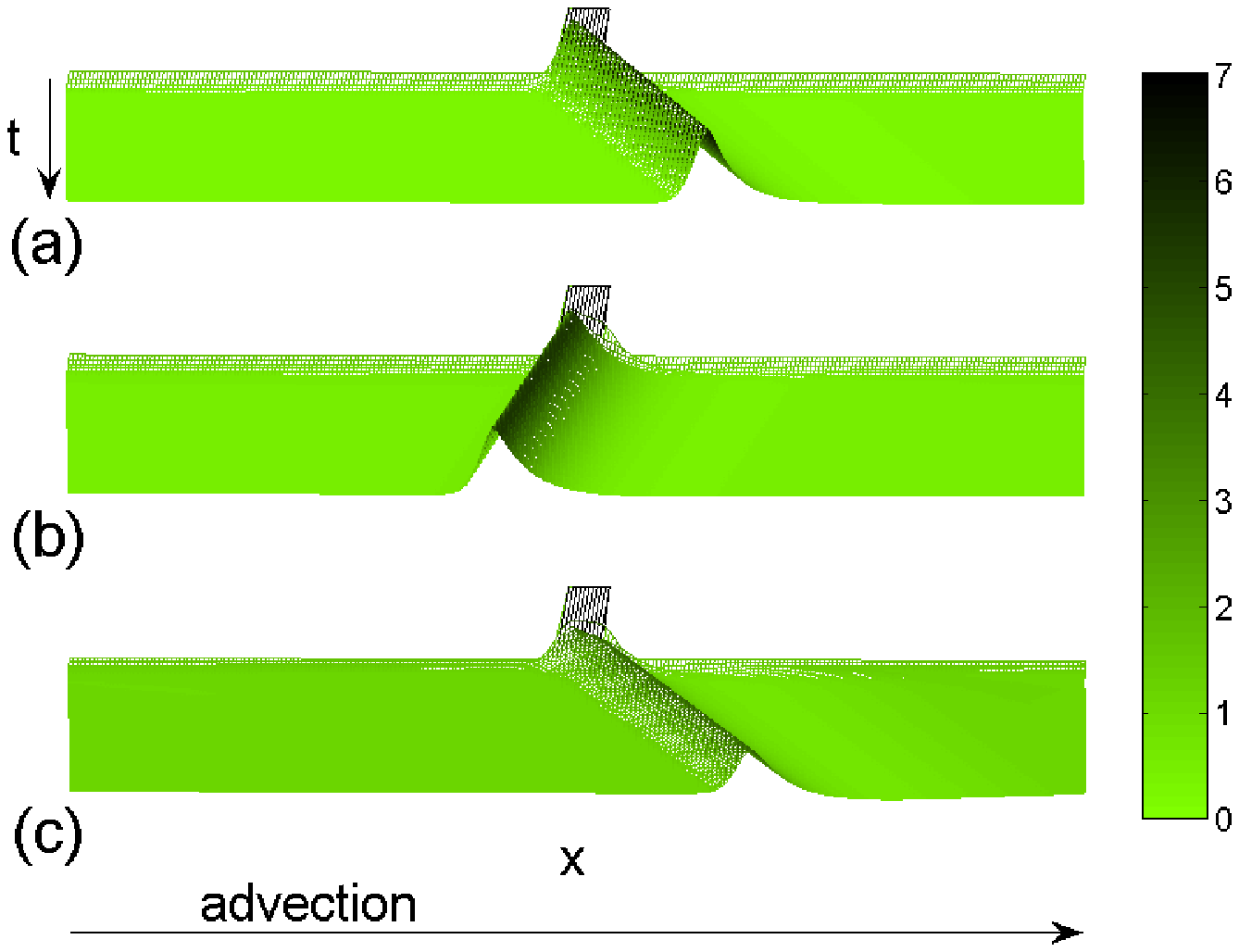}
\caption{(color online) Top panel: Regions of excitable (thick dashed line) and drifting (thin dashed line) solitary waves (pulses) in a
reaction rate parameter space ($B,Da$) at $Le=100$; the thin dashed line implies zero velocity of a pulse. The solid line mark the onsets
of finite wavenumber instabilities of traveling waves, $TW^\pm$. The dotted line marks the criterion for stationary periodic ($SP$)
solutions (see text for details). The ({\large$\bullet$}), marks the leftmost limits of homoclinic orbits $(B^*,Da^*)\simeq(8.76,0.155)$
and asymmetric finite wavenumber Hopf bifurcation $(B^W,Da^W)\simeq(10,0.36)$ while ($\blacksquare$) marks the leftmost limit of excitable
pulses $(B_0,Da^*)\simeq(10.35,0.155)$. Bottom panel: Space-time plots at $B=10.6$ and (a) $Da=0.24$, (b) $Da=0.15$, and (c) $Da=0.08$. The
plots show $v(x,t)$ resulting from integration of Eq.~(\ref{eq:PDE}) with no-flux boundary conditions, where $x\in[0,10]$ and
$t\in[0,2265]$; we used a top-hat initial condition embedded in $(u_0,v_0)$ at the respective $Da$ values.} \label{fig:fig1}
\end{figure}

In this Letter we analyze an RDA case model and demonstrate that under low reaction rates, solitary waves may become convectively unstable
and thus \textit{drift} (see Fig.~\ref{fig:fig1}), i.e. the slow recovery becomes a leading front. We reveal the regions and the properties
of such drifting pulses and show that the phenomenon underlies a competition between a local kinetics of the activator and a differential
advection. Our methods include a bifurcation theory of coexisting spatial solutions (linear analysis and numerical continuations) coupled
to temporal stability; all the results well agree with direct numerical integrations. At the end, we discuss the potential applicability of
our findings to chemical and biological media.

\textit{Model setup.--} We start with a minimal RDA model that incorporates local kinetics of activator $v(x,t)$ and inhibitor $u(x,t)$
type:
\begin{eqnarray}\label{eq:PDE}
      u_t+ s u_x &=& f(u,v)-u, \\
\nonumber  Le\, v_t+ s v_x&=& B f(u,v)-\alpha v +{Pe^{-1}}v_{xx}.
\end{eqnarray}
These dimensionless equations describe a membrane (or cross-flow) reactor, with continuous feeding and cooling in which an exothermic
reaction takes place $f(u,v) \equiv Da (1-u)\exp{[\gamma v/(\gamma+v)]}$~\cite{NRB:00}. Eq.~(\ref{eq:PDE}) admits a uniform rest state
$(u,v)=(u_0,v_0\equiv B u_0/\alpha)$, where $u_0$ obtained via $Da=u_0(1-u_0)^{-1}\exp{[-\gamma u_0/(\gamma \alpha/B+u_0)]}$. In what
follows, we set $Pe=15$, $s=1$, $\alpha=4$, $\gamma=10000$, and use $Le$, $Da$ and $B$ as control parameters allowed to vary; parameter
definitions are given in~\cite{const}.

A standard linear stability analysis to periodic perturbations, shows that the uniform states $(u_0,v_0)$ may loose stability to two finite
wavenumber Hopf instabilities, $Da^\pm$, that emerge from ($B^W,Da^W$), as shown in Fig.~\ref{fig:fig1}; the instabilities are of a
drifting type, i.e. in direction of advection. This anomaly arises due to the broken reflection symmetry of left-right traveling waves that
is preserved in RD systems. We note that traveling waves $TW^-$ bifurcate (nonlinearly) subcritically from $Da^-$ while traveling waves
$TW^+$ bifurcate supercritically from $Da^+$~\cite{YoSh}. While the region $Da^-<Da<Da^+$ is linearly unstable, under certain conditions
stationary periodic (SP) solutions may also develop. The criterion for SP states is zero of the real and the imaginary parts in the
dispersion relation (for a finite wavenumber), identifying zero speed~\cite{YoSh} (see dotted line in Fig.~\ref{fig:fig1}).

Here, our interest is in the affect of a differential advection ($Le$) and the local kinetics ($B,Da$) on the organization of drifting
solitary waves. We also consider large domains in which pulse behavior is not affected by the type of boundary conditions (periodic,
no-flux or mixed) and also not interested in the regimes in which nonuniform steady state patterns may form, for details on the affect of
boundary conditions see~\cite{YoSh}.

\textit{Propagation of solitary waves.--} To reveal the propagation properties and the regimes of solitary waves (see Fig.~\ref{fig:fig1}),
we look at the steady state version of~(\ref{eq:PDE}) in a comoving frame, $\xi=x-ct$~\cite{YoSh}:
\begin{eqnarray}\label{eq:ODE}
&& u_\xi = \bra{s-c}^{-1}\Bra{Da f(u,v)-u}, \quad v_\xi=w, \\
\nonumber  &&w_\xi= Pe\Bra{\bra{s-cLe}w-B Da f(u,v)+\alpha v}.
\end{eqnarray}
The advantage is that existence of nonuniform states can be now analyzed via spatial dynamics methods, i.e. where space is viewed as a
time-like variable. Thus, solitary waves [in the context of~(\ref{eq:PDE})] become in~(\ref{eq:ODE}) asymmetric homoclinic orbits ($HO$)
and $TW^\pm$ (which will be also discussed) correspond to periodic orbits undergoing Hopf bifurcations at $Da^\pm$ (with a proper $c$). In
the following all these solutions will be computed numerically using a continuation package AUTO~\cite{auto}, where the speed $c$ is
obtained as a nonlinear eigenvalue problem. Then temporal stability of such steady states will be calculated employing an eigenvalue
problem via Eq.~(\ref{eq:PDE}) in a comoving frame.
\begin{figure}[tp]
\includegraphics[width=3.1in]{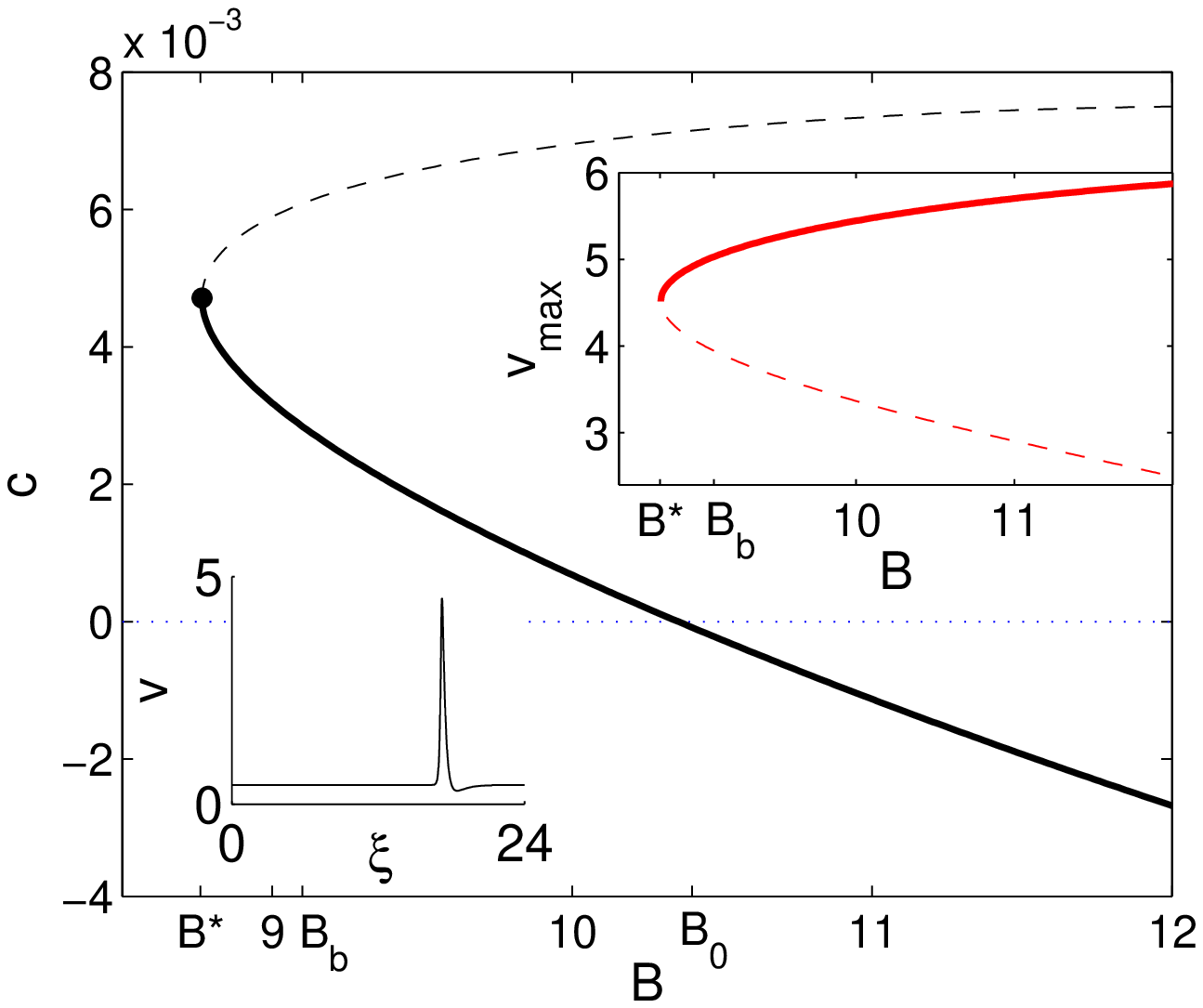}
\includegraphics[width=3.1in]{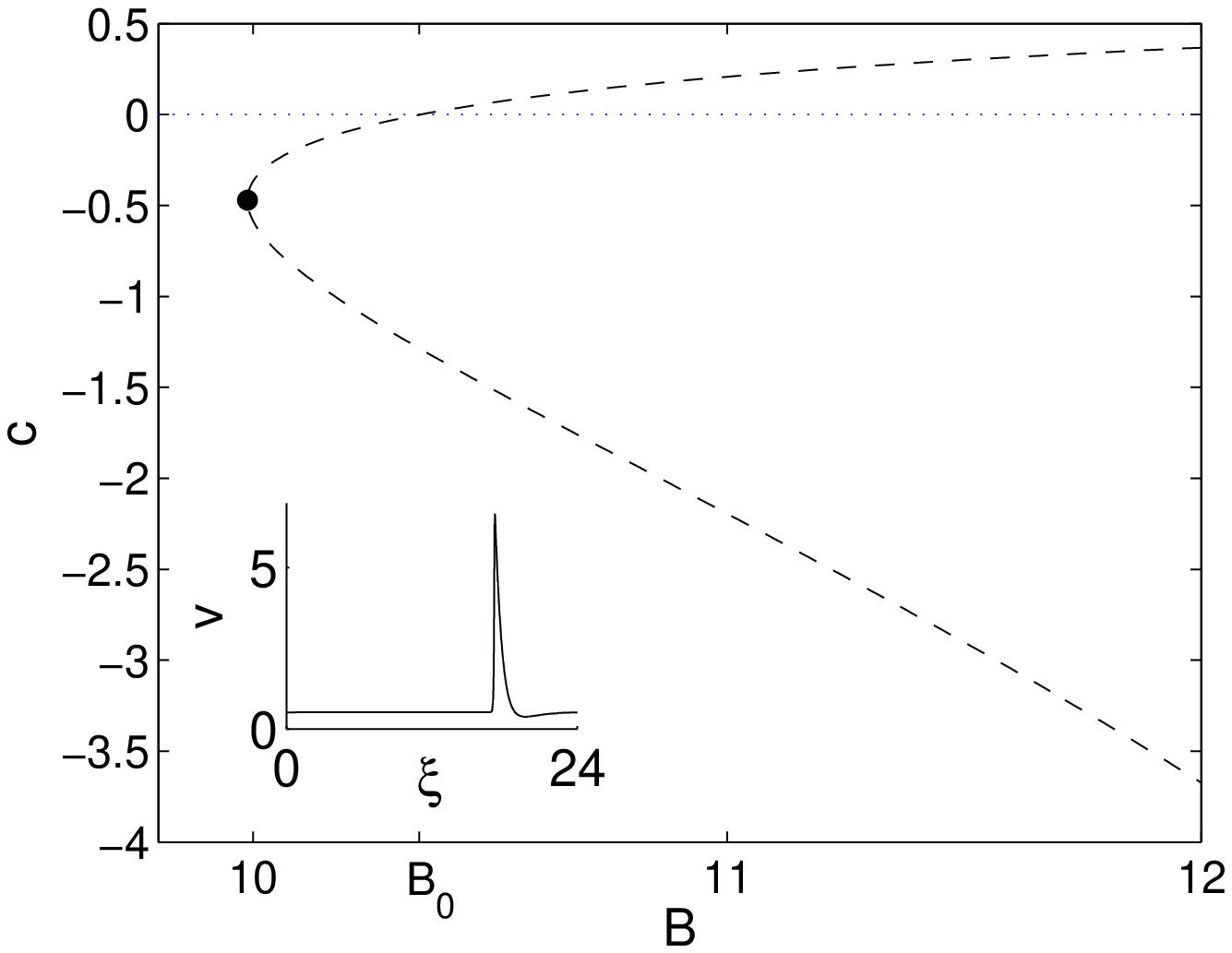} \caption{(color online) Bifurcation diagrams showing
the branches of homoclinic solutions as a function of $B$ at $Da^*\simeq0.155$ and $Le=100$ (top panel) and $Le=1$ (bottom panel). The
branches are plotted in terms of the propagation speed and the maximal value of $v(\xi)$ (top inset); solid lines indicate linear
stability. Bottom insets show profiles of $v(\xi)$ at the two folds, marked by ({\large$\bullet$}). The branches were obtained via
integration of Eq.~(\ref{eq:ODE}) while the stable portions of each branch coincide with solutions obtained by integration of
Eq.~(\ref{eq:PDE}); here the periodic domain is $L=24$ but results are identical also on larger domains.} \label{fig:fig2}
\end{figure}

In Fig.~\ref{fig:fig2}(top panel), we present the branches of $HO$ at $Da=Da^*\simeq 0.155$ (a horizontal cut in top panel in
Fig.~\ref{fig:fig1}), resulting via a simultaneous variation of $(B,c)$. $B=B^*\simeq 8.76$ identifies a fold, where the stable branch
corresponds to large amplitude $HO$ (see inset). The drifting pulses exist for $B^*<B<B_0\simeq 10.35$ since both branches have positive
speeds, and have similar profiles (see inset) as the standard excitable pulses. Namely, drifting pulses propagate in the direction of the
advection (downstream, $c>0$) where the leading front is the oscillatory tail that was a trailing tail above $B=B_0$, for excitable pulses
(upstream, $c<0$).

Drifting pulses are expected at low reaction rate regimes of the activator, represented in~(\ref{eq:PDE}) by dimensionless rate constant
($Da$) and exothermocity ($B$). Under such conditions the excitation of nearest neighbors is suppressed due to the advective flow (a
nonlinear convective instability) and thus the pulse after speed reversal is no longer excitable since the leading front now develops from
the rest state as a small amplitude perturbation. This scenario changes once the differential advection is eliminated ($Le=1$), in this
case a typical RD behavior is restored. While the $c=0$ line for $Le=1$, in the ($B,Da$) plane doesn't change, we show in
Fig.~\ref{fig:fig2}(bottom panel) that near the fold only a negative velocity region forms, i.e. a standard excitable pulses are being
restored (stability of the pulses does not play a qualitative role).

\begin{figure}[tp]
\includegraphics[width=3.3in]{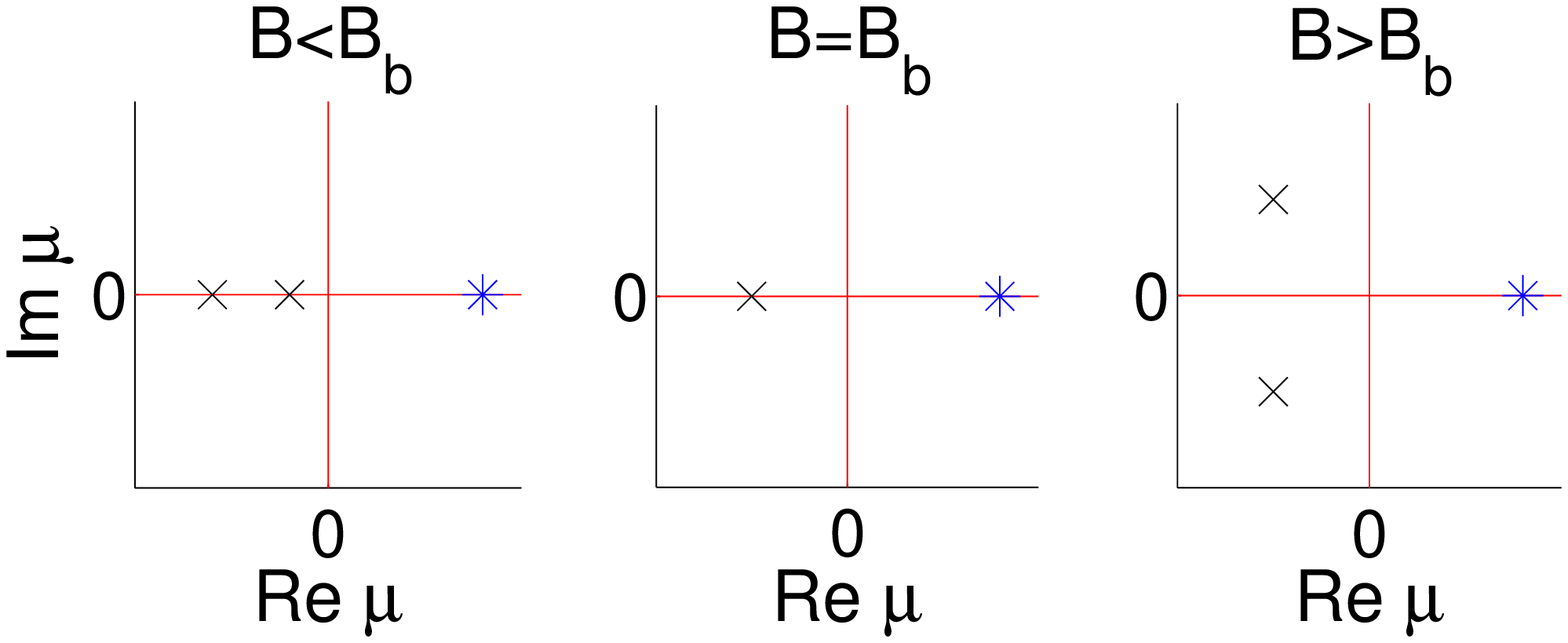}
\includegraphics[width=3.1in]{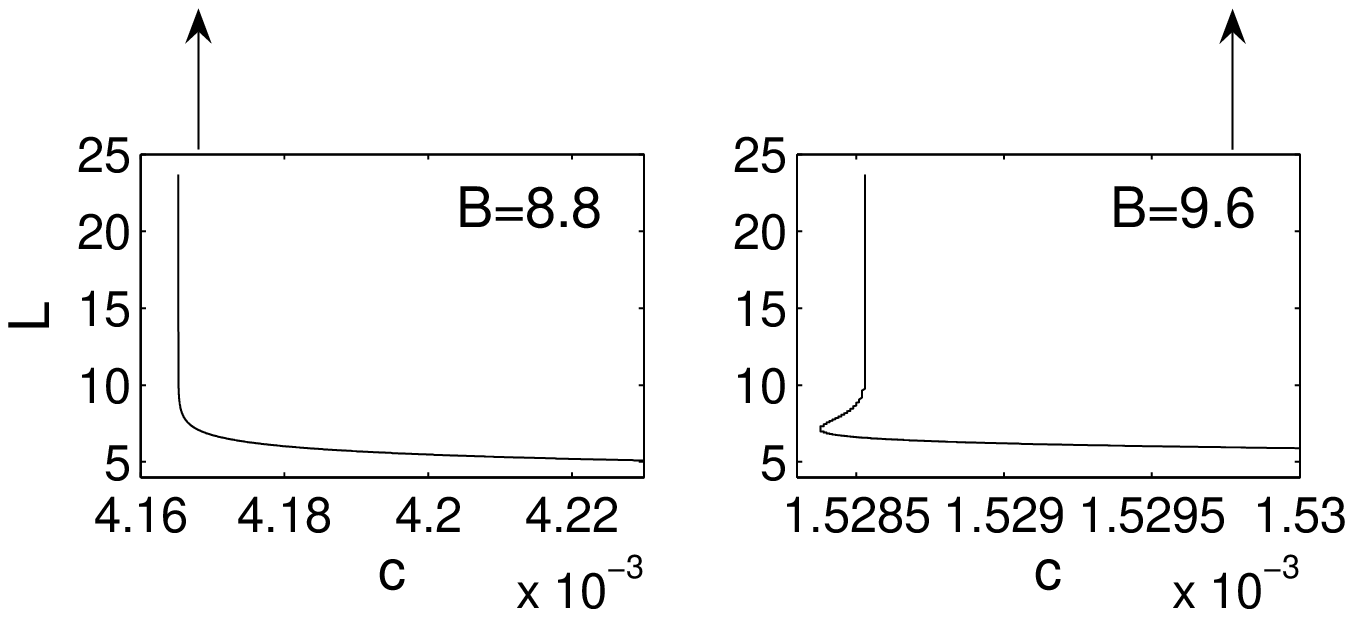}
\caption{(color online) Top panel: Schematic representation of typical eigenvalue configurations about the uniform state $(u_0,v_0,0)$
corresponding to a saddle if $B<B_b$ and a saddle-focus if $B>B_b$, where $B_b \simeq 9.1$ is the Belyakov point. Bottom panel: Typical
dispersion relations that associated with the respective eigenvalues computed starting from the stable homoclinic orbits (see top panel).
Parameters as in the top panel of Fig~\ref{fig:fig2}.} \label{fig:fig3}
\end{figure}

Nevertheless, drifting pulses in presence of a differential advection inherit the properties of excitable pulses, as demonstrated by
monotonic and nonmonotonic dispersion relations in Fig.~\ref{fig:fig3}. The latter are important characteristics of organization and
interaction of solitary waves~\cite{disp_rel} and are distinguished here around $B=B_b\simeq 9.1$, a so called Belyakov
point~\cite{Bel:80}. At this point and with an appropriate speed, the spatial eigenvalues [of Eq.~(\ref{eq:ODE})] correspond to one
positive real (associated with $\xi \to -\infty$) and a degenerate pair of negative reals (associated with $\xi \to \infty$). Below $B_b$,
the degeneracy is removed but the eigenvalues remain negative reals (a saddle) while above $B_b$ they become complex conjugated
corresponding to a saddle-focus (a Shil'nikov type $HO$~\cite{excitable3}), marked by ($\times$) in top panel of Fig.~\ref{fig:fig3}.
Importantly, such an interchange of eigenvalues implies a transition from monotonic to oscillatory dispersion relation
[Fig.~\ref{fig:fig3}(bottom panel)] and a monotonic (in space) approach of the $HO$ to the fixed point as $\xi \to \pm \infty$, which
implies coexistence of bounded-pulse states for $B>B_b$~\cite{disp_rel}.

\begin{figure}[tp]
\includegraphics[width=3.1in]{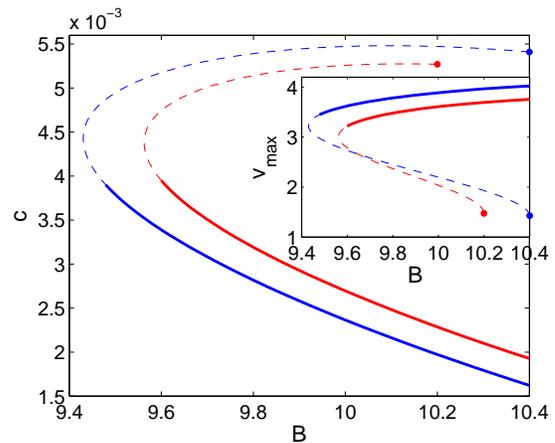}
\caption{(color online) Bifurcation diagram showing the branches of traveling waves ($TW^-$) as a function of $B$ in terms of speed and the
maximal value of $v(\xi)$ (in the inset), at $B=10.4$ (dark line) and $B=10.2$, where $Da^-\simeq 0.29, k_c\simeq 3.2, c\simeq 0.0054$ and
$Da^-\simeq 0.31, k_c\simeq 3.355, c\simeq 0.0053$, respectively. Solid lines imply linear stability to long wave lengths
perturbations~\cite{stab}, while ({\large$\bullet$}) marks the respective onsets of the linear finite wavenumber Hopf bifurcation to
$TW^-$. Integration details as in the top panel of Fig.~\ref{fig:fig2} but on distinct periodic domains.}
\label{fig:fig4}
\end{figure}

\textit{Organization of drifting states.--} A standard theory of solitary waves qualitatively predicts an organization of $HO$ to be
accompanied by periodic solutions~\cite{excitable3}. Here the dispersion relations obtained at $B<B^W$ [Fig.~\ref{fig:fig3}(bottom panel)],
indeed imply existence of periodic orbits although the uniform state is linearly stable. These periodic solutions are in fact $TW^-$ that
bifurcate subcritically from the locus of points $Da=Da^-$ for $B>B^W$ [with distinct critical wavenumbers and speeds obtained from the
linear analysis of Eq.~(\ref{eq:PDE})], as shown by two examples in Fig.~\ref{fig:fig4}. Notably, there are infinite number of such $TW^-$
families. Unlike the $HO$, stability of $TW^-$ solutions do depend on domain size~\cite{stab}.

\begin{figure}[tp]
\includegraphics[width=3.1in]{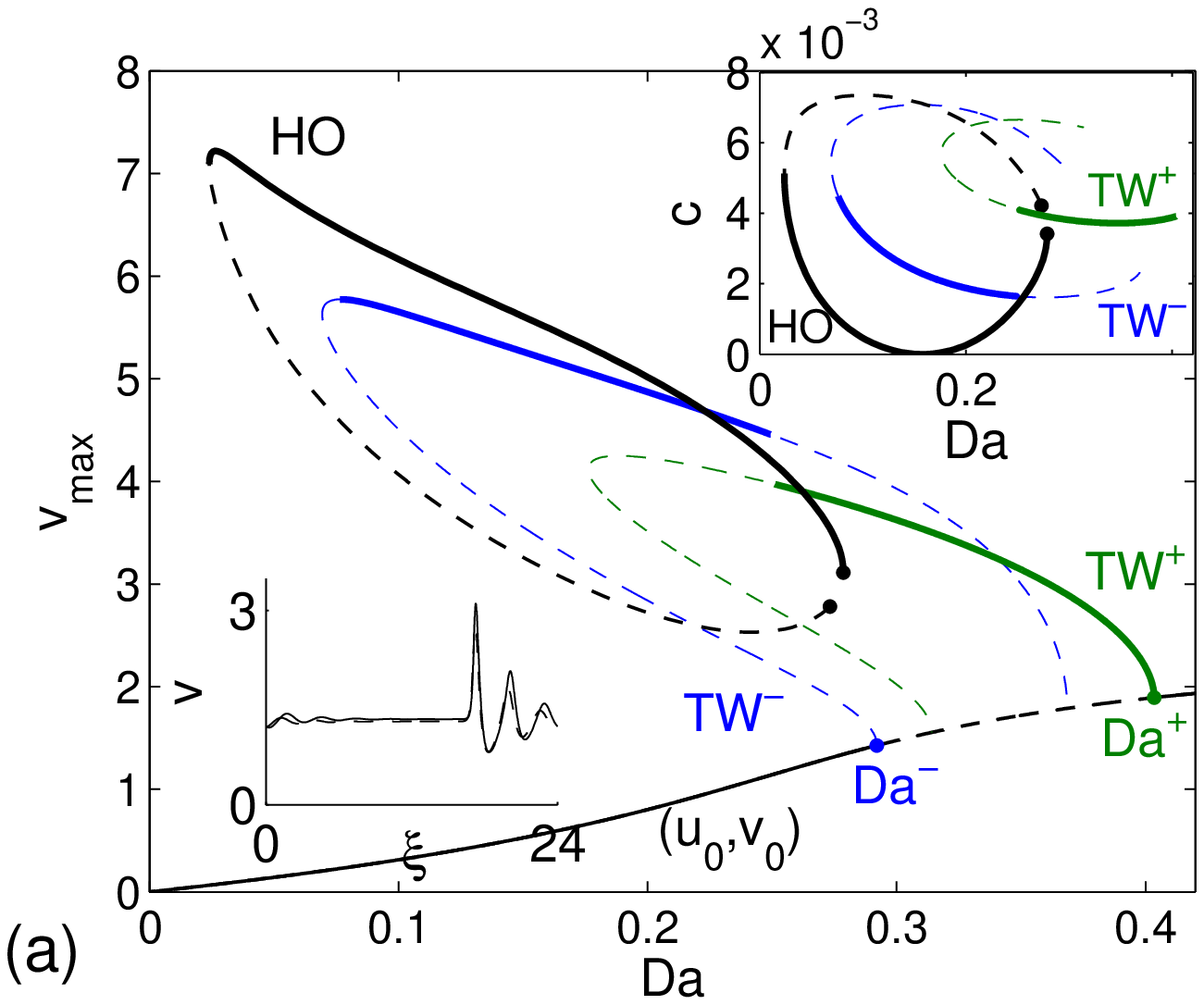}
\includegraphics[width=3.1in]{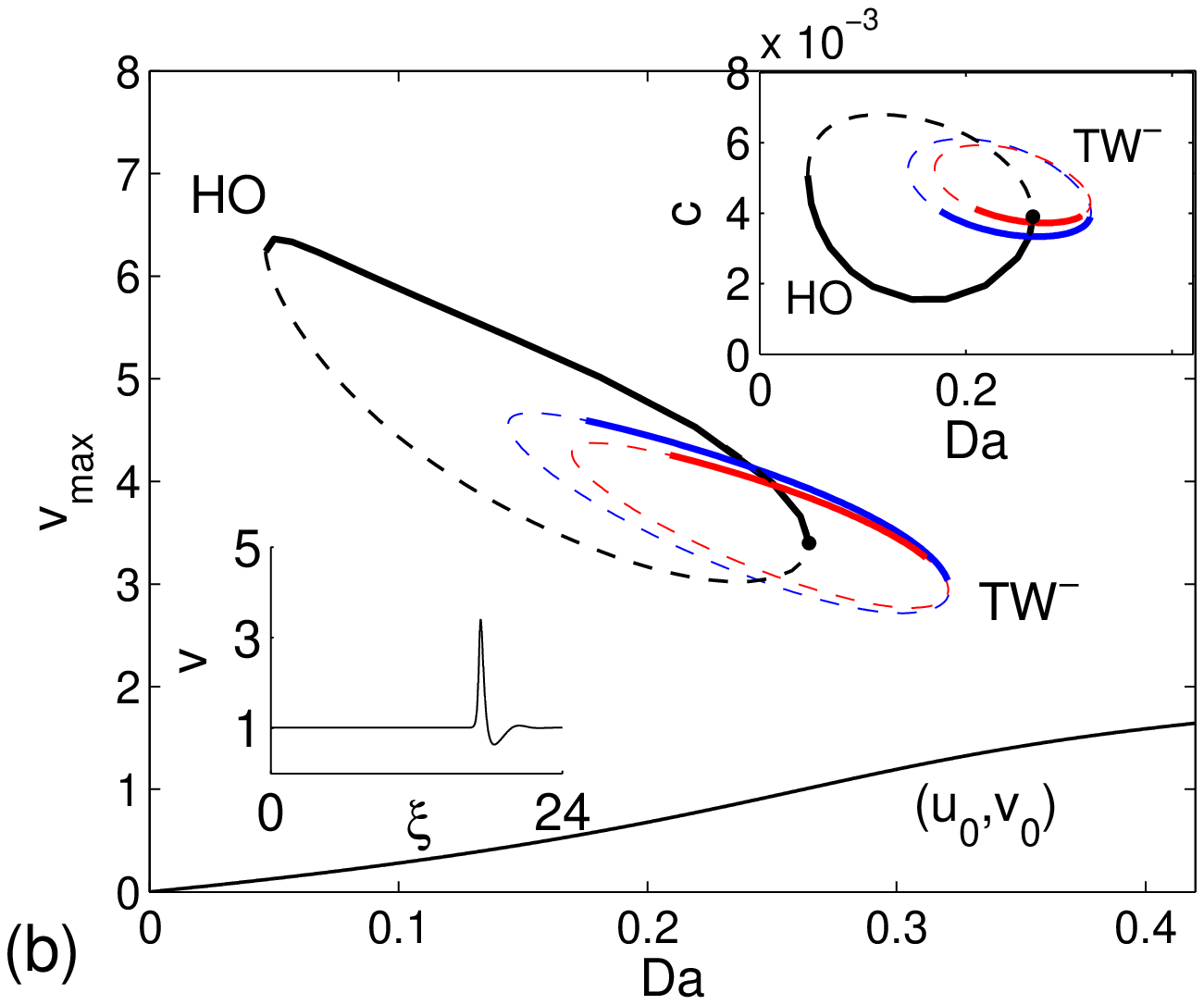}
\caption{(color online) Bifurcation diagram showing the branches of uniform states ($u_0,v_0$), homoclinic orbits ($HO$), and traveling
waves ($TW^\pm$) as a function of $Da$ in terms of the maximal value of $v(\xi)$ at (a) $B=10.4$ and (b) $B=9.6$. Solid lines imply linear
stability, including stability of $TW^\pm$ to long wave lengths perturbations~\cite{stab}, while $Da^\pm$ mark the onsets of the linear
finite wavenumber Hopf bifurcation to $TW^\pm$, respectively. The top inset represents the nonuniform states in terms of speed while the
large (small) isola corresponds to the $TW^-$ family emerging from $Da^-$ at $B=10.4$ ($B=10.2$). The bottom shows $HO$ profiles at
locations marked by ({\large$\bullet$}); in (a) the two dots mark also the two ends of the $HO$ branch. Integration details as in the top
panel of Fig.~\ref{fig:fig2} but on distinct periodic domains.}
\label{fig:fig5}
\end{figure}

The organization of all drifting nonuniform solutions can be understood by varying $Da$ at two representative $B$ values.
Fig.~\ref{fig:fig5}(a), shows a bifurcation diagram of nonuniform solutions at $B\simeq 10.4$: while $TW^\pm$ propagate downstream. The
single pulse $HO$ branch ends at the two rightmost ends (marked by dots), at which the profiles take the form of homoclinic tails (see
bottom inset)~\cite{SLY:00}. Due to the proximity to the subcritical onset of $TW^-$ at $Da^-$, the two rightmost ends ever approach each
other as domain ($L$) is increased, and consequently, they inherit the propagation direction of the top and the bottom branches of $TW^-$
as discussed in~\cite{YoSh}. As $B$ is decreased below $B^W$ the $HO$ and the $TW^-$ solutions organize in isolas and parts of their
stability regions overlap [Fig.~\ref{fig:fig5}(b)], implying sensitivity to initial perturbations. Note that the oscillations of the right
tail in the profile had decreased (see bottom inset), which is consistent with the approach towards Belyakov point ($B=B_b$).

\textit{Conclusions and prospects.--} We have showed that solitary waves can propagate bidirectionally (without changing their shape) due
to a competition between activator autocatalysis and a symmetry breaking advection. Consequently, we distinguish between excitable
(upstream or against advection) and drifting (downstream or with advection) propagations. The former is a characteristic behavior of RD
systems and persists while the reaction rate of the activator is dominant (analogues to front dynamics~\cite{front}). While the latter is a
consequence of low excitation and thus subjected to a nonlinear convective instability resulting in a fluid type behavior. Through a
bifurcation analysis of spatial extended steady states arising in a minimal RDA model, we revealed the properties and the organization of
drifting pulses. Since the results center on homoclinic orbits which known to act as organizing centers of spatial solutions, qualitative
applicability to systems with other autocatalytic properties is naturally anticipated.

Up-to-date only excitable (upstream) solitary waves have been observed experimentally in an autocatalytic RDA system~\cite{KM:02},
nevertheless chemical media operated in cross-flow (membrane) tubular reactors~\cite{NRB:00} or on a rotating disks~\cite{KhPi:95}, are the
most natural setups to confirm our predictions and explore technological directions. Moreover, theoretical insights explored here can be
related to a profound puzzle of large intracellular particles (organelles) self-organization, in eucaryotic cells~\cite{We:04}. For
example, localized aggregations of myosin-X within the filopodia have been observed to propagate bidirectionally~\cite{BeCh:02} and from
the modeling point of view argued to be driven by both diffusion and differential advection~\cite{motors}. Consequently, a theoretical
framework integrating autocatalytic kinetics and distinct transport, is paramount to promoting a mechanistic understanding of
spatiotemporal trafficking of intracellular molecular aggregations.

We thank to N. Gov and M. Naoz for the helpful discussion on molecular motors. This work was supported by the US-Israel Binational Science
Foundation (BSF) and A.Y. was also partially supported by the Center for Absorption in Science, Israeli Ministry of Immigrant Absorbtion.
M.S. is a member of the Minerva Center of Nonlinear Dynamics and Complex Systems.



\begin{thebibliography}{30}
\bibitem{const}
$Le$ (Lewis number), is the ratio of solid- to fluid-phase heat capacities, $Pe$ (P\'{e}clet number), is the ratio of convective to
conductive enthalpy fluxes, and $Da$ (Damk\"{o}hler number), is the dimensionless rate constant~\cite{NRB:00}.
\bibitem{stab}
Temporal stability of $TW^\pm$ was computed for large periodic domains, $L=n\lambda_c, n>1,\lambda_c\equiv 2\pi/k_c$ (until the onset
didn't change with $n$), via a standard numerical eigenvalue method using Eq.~(\ref{eq:PDE}) in a comoving frame. $k_c$ is the critical
wavenumber at the Hopf onset.
\end{thebibliography}
\end{document}